\documentclass[twocolumn,runningheads,natbib]{svjour2}

\smartqed  
\usepackage{graphicx}

\newcommand{\B}{\ensuremath{\vec{B}}}         

\newcommand{\GJ}[1]%
{\ensuremath{#1_\textrm{\tiny{}GJ}}} 
\newcommand{\Rotat}[1]%
{\ensuremath{#1^\textrm{\tiny{}rot}}} 
\newcommand{\Osc}[1]%
{\ensuremath{#1^\textrm{\tiny{}osc}}} 
\newcommand{\NS}[1]%
{\ensuremath{#1_\textrm{\tiny{}NS}}} 

\newcommand{\Rgj}{\GJ{\rho}}        
\newcommand{\egj}{\GJ{E}}           
\newcommand{\Pgj}{\GJ{\Psi}}        

\newcommand{\RS}{\NS{R}}                        

\newcommand{\Vosc}{\Osc{V}}
\newcommand{\Vrot}{\Rotat{V}}

\newcommand{\n}{\nabla}  
\newcommand{\D}{\Delta}              
\newcommand{\x}{\times}     
\newcommand{\pd}{\partial}           
\newcommand{\ROT}{\nabla\times}
\newcommand{\DIV}{\nabla\cdot}

\newcommand{\N}{\ensuremath{\vec{e_r}}}                  

\graphicspath{{./figures/}}

\journalname{Astrophysics and Space Science}
\begin{document}

\title{Impact of neutron star oscillations on the accelerating
  electric field in the polar cap of pulsar
}
\subtitle{or could we see oscillations of the neutron star after the glitch
  in pulsar?}

\titlerunning{Could we see oscillations of the neutron star?}        

\author{A.~N.~Timokhin}


\institute{A.~N.~Timokhin\at
              Sternberg Astronomical Institute \\
              Universitetskij pr. 13 \\
              119992 Moscow, Russia
              \email{atim@sai.msu.ru}           
}

\date{Received: date / Accepted: date}

\maketitle

\begin{abstract}
  Pulsar "standard model", that considers a pulsar as a rotating
  magnetized conducting sphere surrounded by plasma, is generalized to
  the case of oscillating star. We developed an algorithm for
  calculation of the Goldreich-Julian charge density for this case.
  We consider distortion of the accelerating zone in the polar cap of
  pulsar by neutron star oscillations.  It is shown that for
  oscillation modes with high harmonic numbers $(l,m)$ changes in the
  Goldreich-Julian charge density caused by pulsations of neutron star
  could lead to significant altering of an accelerating electric field
  in the polar cap of pulsar.  In the moderately optimistic scenario,
  that assumes excitation of the neutron star oscillations by
  glitches, it could be possible to detect altering of the pulsar
  radioemission due to modulation of the accelerating field.
\end{abstract}

\section{Introduction}
\label{sec:introduction}

Neutron stars (NS) are probably the most dense objects in the
Universe.  There are extreme physical conditions inside NS, i.e. the
magnetic field is close to the quantum limit, the pressure is of the
order of the nuclear one and the typical radius of a NS is only about
2-3 times larger than its gravitational radius.  Knowledge of
properties of matter under such extreme conditions if very important
for fundamental physics.  It is impossible to reconstruct such
physical circumstances in terrestrial laboratories, therefore, study
of the NS's internal structure would give an unique opportunity for
experimental verification of several fundamental physical theories.

To study interiors of a celestial body one have to perform some kind
of seismological study, by comparing observed frequencies of
eigenmodes with frequencies inferred from theoretical considerations.
Eigenfrequencies of oscillations in the crust of NS as well as in its
interiors were calculated in several papers \citep[see e.g.  ][and
references there]{McDermott/1988,Chugunov2006}.  However, for
seismological study there should exist both i) a mechanism for
excitation of oscillations, ii) a mechanism modulating radiation of the
celestial object.

There are two types of known NSs: member of binary systems and
isolated ones.  The former radiate due to accretion of the matter from
the companion.  For these stars there are many possibilities to excite
oscillations, for example by instabilities in the accretion flow.
However, in this case it would be difficult to distinguish whether a
particular feature in the power spectrum of the object is due to
oscillations of the NS or it is caused by some processes in the
accretion disc/column.  Because of this ambiguity we think that the
study of isolated NS should be more promising in regard of the
seismology.

The vast majority of known isolated neutron stars are radiopulsars.
The glitch (sudden change of the rotational period) is probably the
only possible mechanism for excitation of oscillations for isolated
pulsars.  Radiation of radiopulsars is produced mostly in the
magnetosphere.  In order to judge whether oscillations of the NS could
produce detectable changes in pulsar radiation, the impact of the
oscillations on the magnetosphere must be considered.  There is a
widely accepted model of radiopulsar as a highly magnetized NS
surrounded by non-neutral plasma \citep{GJ}.  Although, there is still
no self-consistent theory of radiopulsars, there is a general
agreement regarding basic picture for the processes in the
magnetosphere.  Oscillations of the star can generate electric field
as it happens in the case of rotation.  Generalization of the
formalism developed for rotating NS to the case of oscillating star
should help to obtain the desired information.

The first attempt to generalize Goldreich-Julian (GJ) formalism to the
case of oscillating NS was made in \citet{TBS2000}.  It was developed
a general algorithm for calculation of the GJ charge density in the
near zone of an oscillating NS.  Using this algorithm GJ charge
density and electromagnetic energy losses were calculated for the case
of toroidal oscillations of the NS.  Here we apply this formalism to
the case of spheroidal oscillation modes, representing wide class of
stellar modes (r-,g-,p- modes).  We consider also impact of stellar
oscillations on the acceleration mechanism in the polar cap of pulsar
and discuss the possibility of observation on the NS oscillations.

\section{Main Results}

\subsection{General formalism}
\label{sec:general-formalism}

Let us start by considering the case of a non-rotating oscillating NS.
Motion of the conducting NS surface in the strong magnetic field of
the star generates electric field as in the case of rotation.  Only
oscillation modes with non-vanishing velocity $\Osc{\vec{V}}$ at the
surface will disturb the magnetosphere.  For the same reason as it is
in the pulsar ``standard model'', the electric field in the
magnetosphere of an oscillating star should be perpendicular to the
magnetic field.  Otherwise charged particles will be accelerated by a
longitudinal (parallel to \B) electric field and their radiation will
give rise to electron-positron cascades producing enough particles to
screen the accelerating electric field \citep{Sturrock71}.  As in the
case of rotating stars we will define the Goldreich-Julian electric
field $\GJ{\vec{E}}$ as the field which is everywhere perpendicular to
the magnetic field of the star, $\GJ{\vec{E}}\perp\B$, and the GJ
charge density as a charge density, which supports this field
\begin{equation}
  \label{eq:Rgj}
  \Rgj \equiv \frac{1}{4\pi} \nabla\cdot\GJ{\vec{E}}\:.
\end{equation}

For simplicity we consider only a zone near the NS, at the distances
$r\ll{}2\pi c/\omega$, where $\omega$ is the
frequency of NS oscillations.  For many global oscillation modes
\citep[see e.g.][]{McDermott/1988} the polar cap accelerating zone is
well withing this distance, therefore, we can study the changes of the
accelerating electric field in the polar cap caused by oscillation.
In the near zone all physical quantities change harmonically with
time, i.e. the time dependence enters only through the term
$e^{-\mathrm{i}\omega t}$.

We make an additional assumption, that changes of the magnetic field
induced by currents in the NS crust are much larger than the
distortion caused by currents flowing in the magnetosphere.  This
assumption is considered as a first order approximation according to
the small parameter $\left(\xi/\RS\right)$, where $\xi$ is the
amplitude of oscillation and \RS -- the NS radius. In other words,
outside of the NS 
\begin{equation}
\ROT \B = 0\:
\label{rot_B_eq_0}
\end{equation}
in the first order in $\left(\xi/\RS\right)$.  This assumption can be
rewritten in terms of a condition on the current density in the
magnetosphere as
\begin{equation}
  j \ll \Rgj \; c\; \left(\frac{c}{\omega r}\right)
  \equiv \GJ{j} \left(\frac{c}{\omega r}\right)\:.
\label{eq:j_ll_B_Rgj}
\end{equation}
Condition~(\ref{eq:j_ll_B_Rgj}) implies that the current density in
the near zone of the magnetosphere is less that the GJ current density
connected with oscillations multiplied by a large factor $c/(\omega
r)$.  So, if the current density in the magnetosphere is of the order
of the GJ current density, assumption~(\ref{rot_B_eq_0}) is valid.
Under these assumption it is possible to solve the problem
analytically in general case, i.e. to develop an algorithm for finding
an analytical solution for the GJ charge density for arbitrary
configuration of the magnetic field and arbitrary velocity field on
the NS surface.

Under assumption (\ref{rot_B_eq_0}) the magnetic field can be
expressed through a scalar function $P$ as
\begin{equation}
  \B = \n \x \n \x (P \N)\: .
  \label{eq:B_P}
\end{equation}
The GJ electric field depends also on a scalar potential $\Pgj$
through the relation
\begin{equation}
  \GJ{\vec{E}} = - \frac{1}{c} \n \x (\pd_t P \N) - \n \Pgj \: .
\label{eq:E_P_Psi}  
\end{equation}
The GJ charge density is then expressed as
\begin{equation}
  \label{eq:Rgj_Psi} 
  \Rgj = - \frac{1}{4\pi} \D \Pgj\: .
\end{equation}
An equation for $\Pgj$ is
\begin{eqnarray}
  \label{eq:EquationGeneral}
  \D_{\Omega} P \; \pd_r\Pgj -
  \pd_r \pd_\theta P \; \pd_\theta \Pgj -
  \frac{1}{\sin^2\theta}\: \pd_r \pd_\phi P \; \pd_\phi \Pgj \nonumber  \\ 
  + \frac{1}{c \sin\theta}
  \left( \pd_r \pd_\phi P \; \pd_\theta \pd_t P -
    \pd_r \pd_\theta P \; \pd_\phi \pd_t P
  \right)  =  0\:,
\end{eqnarray}
where $\D_{\Omega}$ is an angular part of the Laplace operator.  This
is the first order linear partial differential equation for the GJ
electric potential \Pgj.  As the equation for \Pgj{} is linear, each
oscillation mode can be treated separately.  This equation is valid
for arbitrary configuration of the magnetic field and for any
amplitude of the surface oscillation, provided that condition
(\ref{rot_B_eq_0}) is satisfied.  Dependence on oscillation mode
appears in the boundary conditions and also through the time
derivative $\pd_t P$.  For different oscillation modes both the
equation and boundary conditions for \Pgj{} are different.  Derivation
of eq.~(\ref{eq:EquationGeneral}), boundary condition for functions
$\Pgj$ as well as more detailed discussion of used approximations can
be found in \citet{TBS2000}.

\begin{figure*}
  \centering
  \includegraphics[width=\columnwidth]{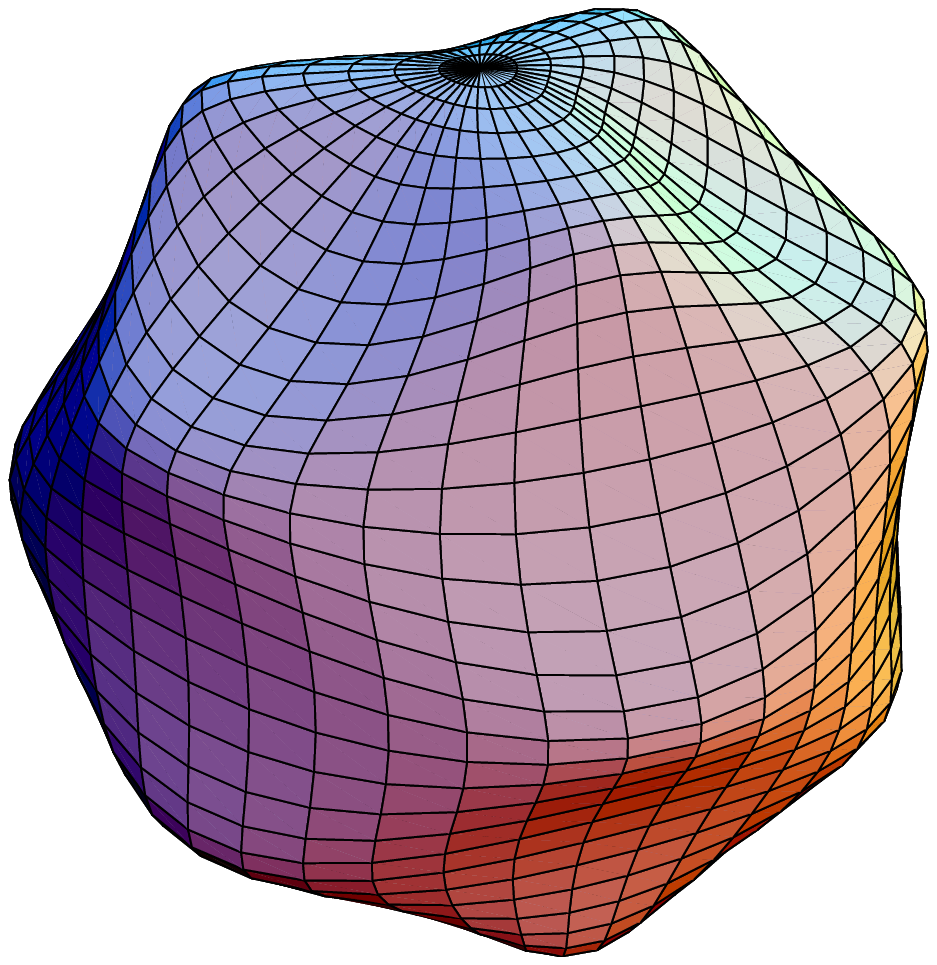}
  \includegraphics[width=\columnwidth]{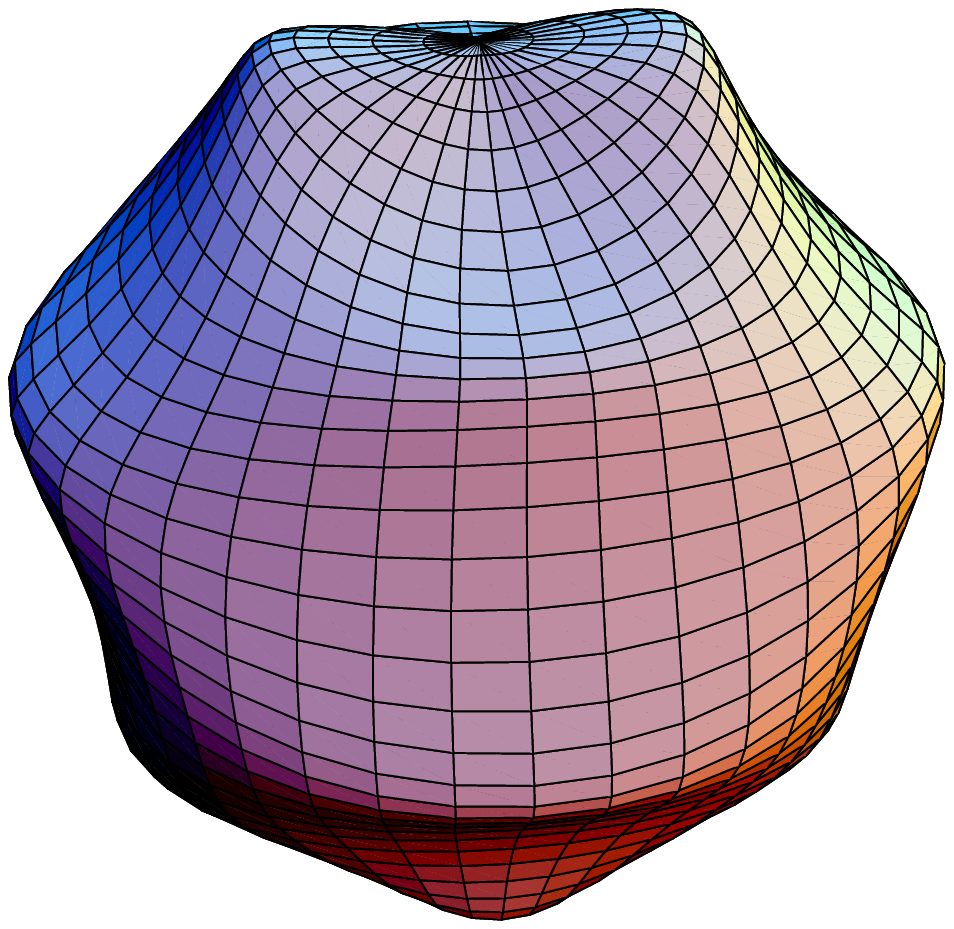}
  \caption{The shape of the star surface during oscillations
    \textit{left} for spheroidal mode $(7,3)$, \textit{right} for
    spheroidal mode $(7,2)$. The shape changes harmonically with time.}
\label{fig:shape}
\end{figure*}

\begin{figure*}
  \centering
  \includegraphics[width=.95\columnwidth]{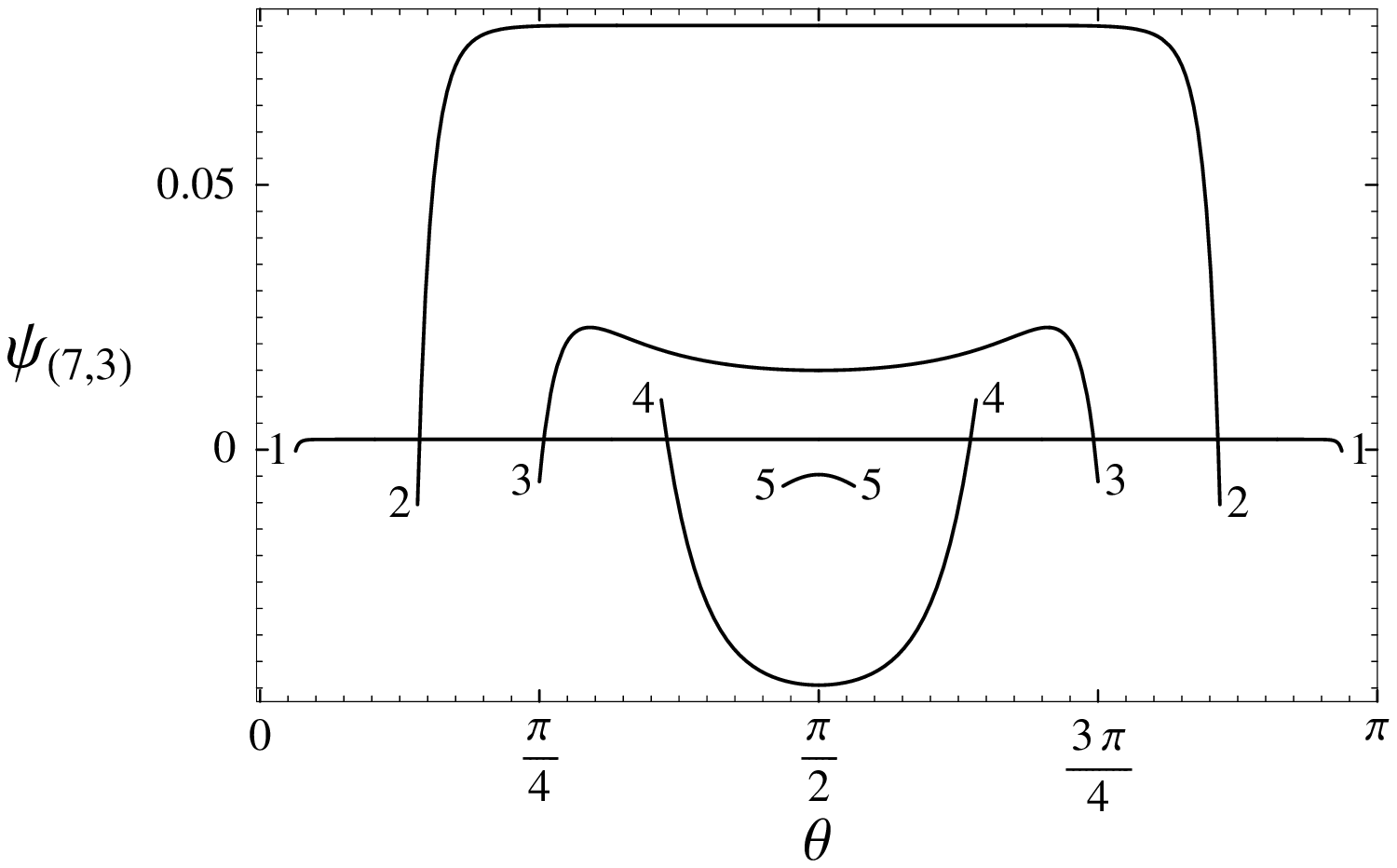}
  \hspace{0.5cm}
  \includegraphics[width=.95\columnwidth]{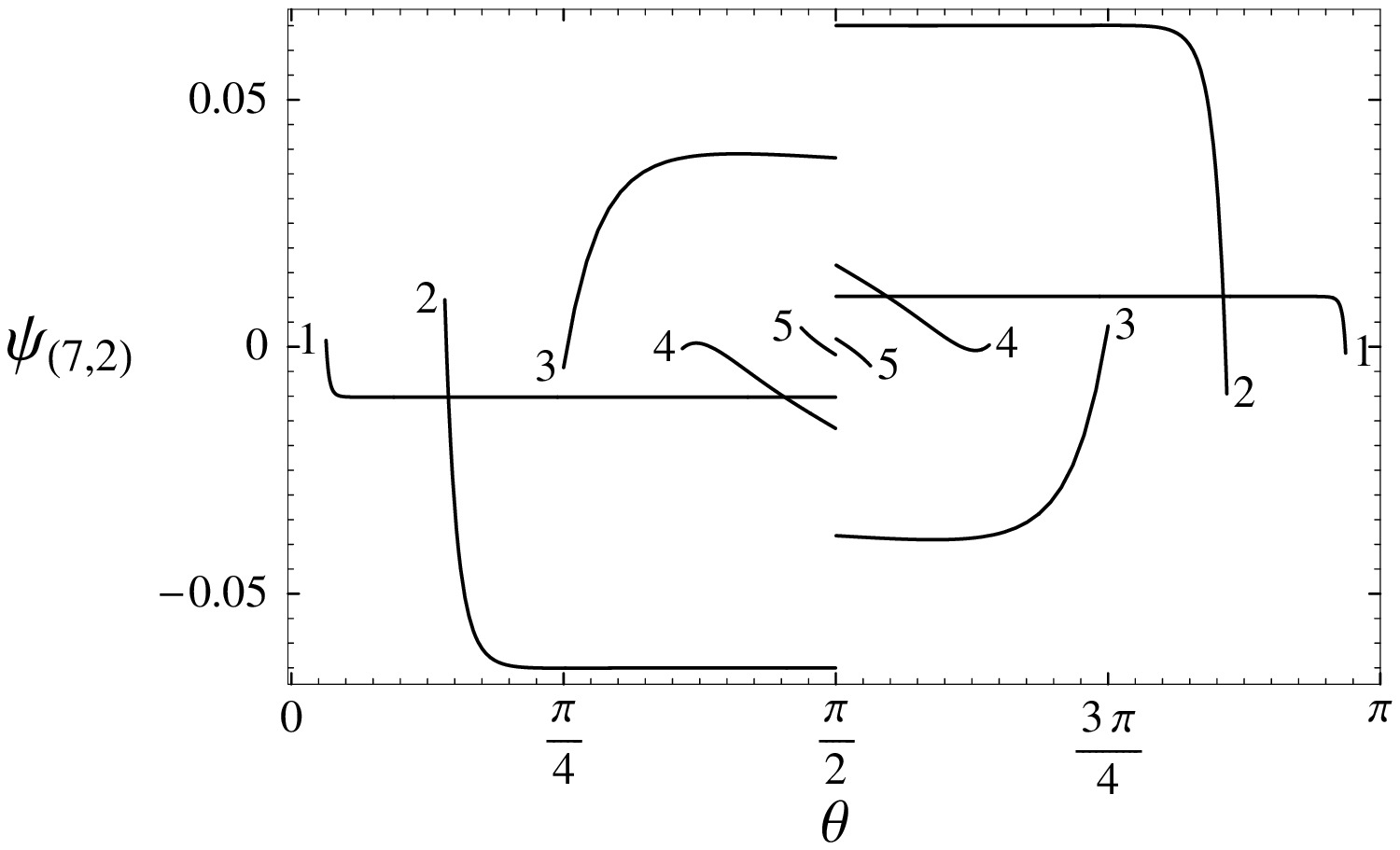}
  \caption{Electric potential $\Pgj$ along a dipolar magnetic field
    line as a function of the polar angle $\theta$ is shown for 5
    field lines with azimuthal angle $\phi=0$.  \textit{left} for
    spheroidal mode $(7,3)$, \textit{right} for spheroidal mode
    $(7,2)$. The quantity changes harmonically with time.}
\label{fig:Pgj}
\end{figure*}

\begin{figure*}
  \centering
  \includegraphics[width=0.95\columnwidth]{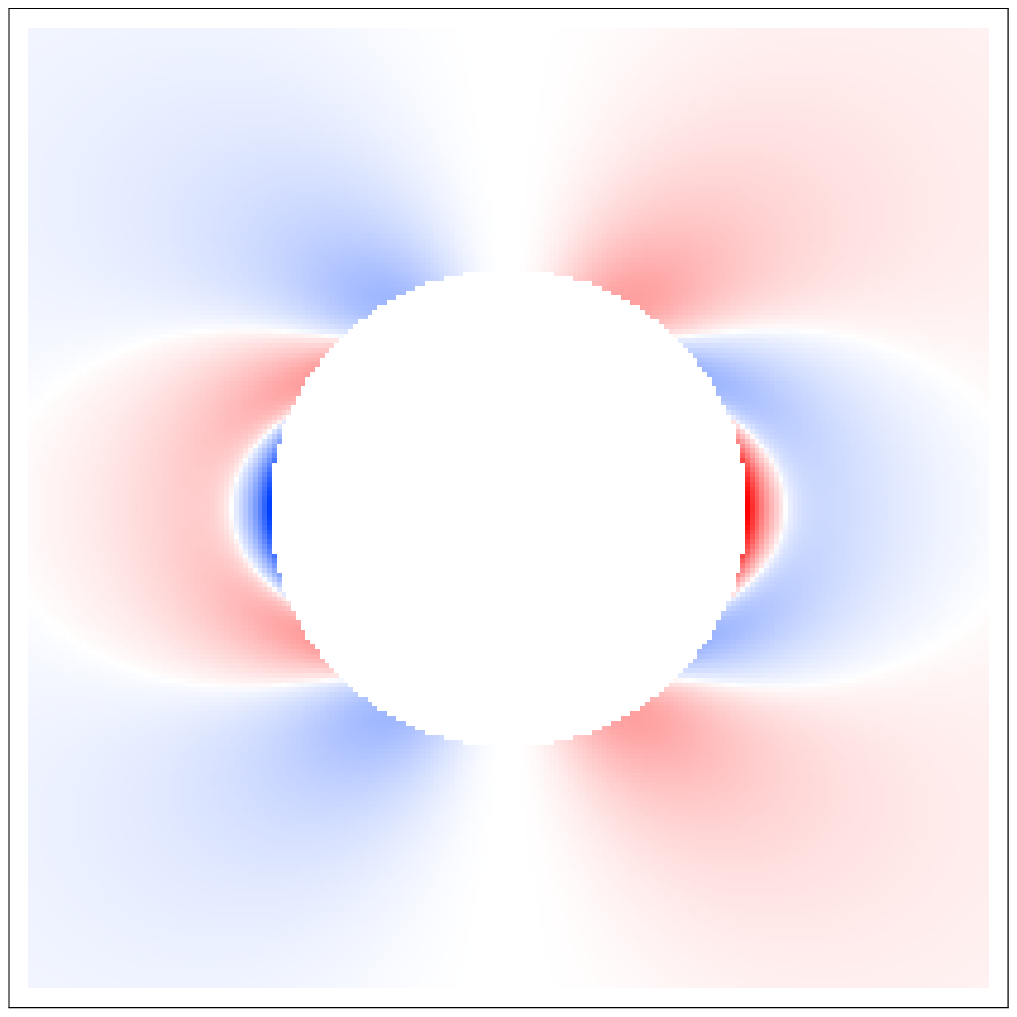}
  \hspace{0.5cm}
  \includegraphics[width=0.95\columnwidth]{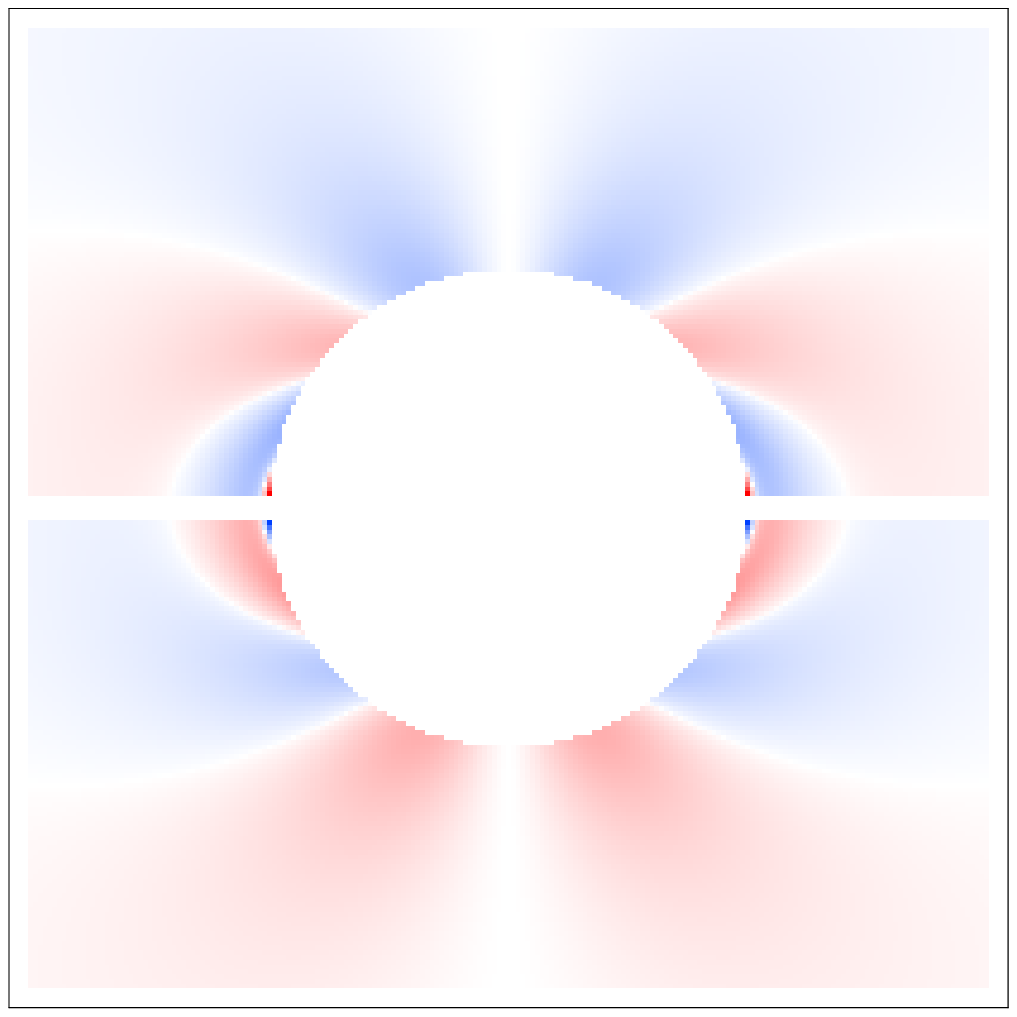}
  \caption{Charge density $\Rgj$ near the NS for azimuthal angle
    $\phi=0$. Positive values of the charge density are shown in red
    and negative ones in blue.  \textit{left}: $\Rgj$ for the
    spheroidal mode $(7,3)$; \textit{right}: $\Rgj$ for the spheroidal
    mode $(7,2)$. The quantity changes harmonically with time. NB: at
    the equatorial plane ($\theta=\pi/2$) $\Rgj^{72}$ is infinite. }
\label{fig:Rgj_map}
\end{figure*}

\begin{figure*}
  \centering
  \includegraphics[width=0.95\columnwidth]{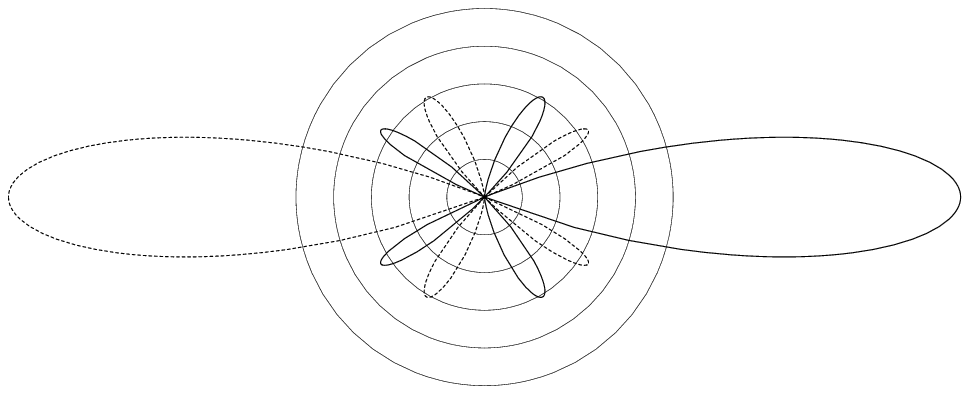}
  \hspace{0.5cm}
  \includegraphics[width=0.95\columnwidth]{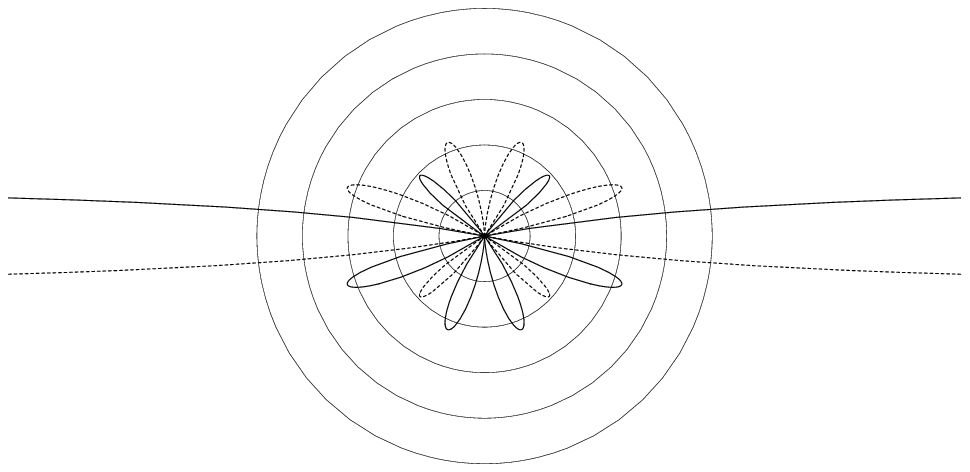}
  \caption{Charge density $\Rgj(r=\RS,\theta,\phi=0)$ on the NS surface
    is shown in a polar coordinate system
    $(\left|\Rgj^{lm}(\RS,\theta,0)\right|, \theta)$.  Positive values
    of charge density are shown by the solid line and negative ones by
    the dashed line.  \textit{left}: $\Rgj$ for the spheroidal mode
    $(7,3)$; \textit{right}: $\Rgj$ for the spheroidal mode $(7,2)$.
    The quantity changes harmonically with time. Circles correspond to
    the values of $\Rgj=(0.1,0.2,0.3,0.4,0.5)$ in normalized units.
    NB: on the equatorial plane ($\theta=\pi/2$) $\Rgj^{72}$ is
    infinite. }
\label{fig:Rgj_surf}
\end{figure*}

\subsection{Goldreich Julian charge density}
\label{sec:goldr-juli-charge}

In \citet{TBS2000} equation~(\ref{eq:EquationGeneral}) was solved for
the case of small-amplitude toroidal oscillations and dipolar
configuration of unperturbed magnetic field.  Solutions had been
obtained with a code written in computer algebra language MATHEMATICA.
Now we have developed a new version of this code, which allows to
obtain analytical solutions of equation~(\ref{eq:EquationGeneral}) for
a more complicated case of spheroidal modes.  Any vector field on a
sphere can be represented as a composition of toroidal
($\DIV\Osc{\vec{V}} = 0$) and spheroidal ($\ROT\Osc{\vec{V}} = 0$)
vector fields \citep{Unno1979}.  So, now we are able to calculate GJ
electric field and charge density for arbitrary oscillations of a NS
with dipole magnetic field.

Similar to the case of toroidal oscillations the small current
approximation turned to be valid for a half of all oscillation modes.
For oscillation modes with velocity field, which is symmetric relative
to the equatorial plane (see an example of such mode in
Fig.~\ref{fig:shape}~(left)), solution of
eq.~(\ref{eq:EquationGeneral}) is smooth everywhere (see
Fig.~\ref{fig:Pgj}~(left)).  For modes with antisymmetrical velocity
field (an example of such mode is shown in
Fig.~\ref{fig:shape}~(right)), \Pgj{} is discontinuous at the
equatorial plane (see Fig.\ref{fig:Pgj}~(right)).  There is the
following reason for such behavior.  Dipolar magnetic field is
antisymmetric relative to the equatorial plane. Antisymmetric motion
of the field line footpoints give rise to a twisted configuration of
the magnetic field, which cannot be curl-free.  So, for such modes a
strong current will flow along closed magnetic field lines,
$j\gg\GJ{j}$.  However, there is no physical reason why a smooth
solution for GJ electric field can not exists also for such modes.  An
argument supporting this hypothesis is a solution for twisted
force-free magnetic field found by \citet{Wolfson1995}.  In his
solution, which corresponds to the toroidal mode $(2,0)$, the
configuration of force-free twisted magnetic field is supported by a
strong current flowing along magnetic field lines.

For the modes with smooth solutions our approximation should be valid.
On the other hand, a strong electric current will flow only along
closed magnetic field lines.  The current density along open magnetic
field lines should be close to $\GJ{j}$ and condition for the small
current approximation (eq.~(\ref{eq:j_ll_B_Rgj})), will be satisfied in
the open field line domain.  Then, \Pgj{} in the polar cap could be
obtained by solving eq.~(\ref{eq:EquationGeneral}).  \Pgj{} in the
polar cap will differ from the solutions obtained here, because the
boundary conditions should be set at the polar cap boundaries and not
on the whole surface on the NS.  The main properties of \Pgj{}
regarding its qualitative dependence on the coordinates, used in our
discussion, would be, however, similar to the properties obtained from
our solutions.

\begin{figure}
  \centering
  \includegraphics[width=.95\columnwidth]{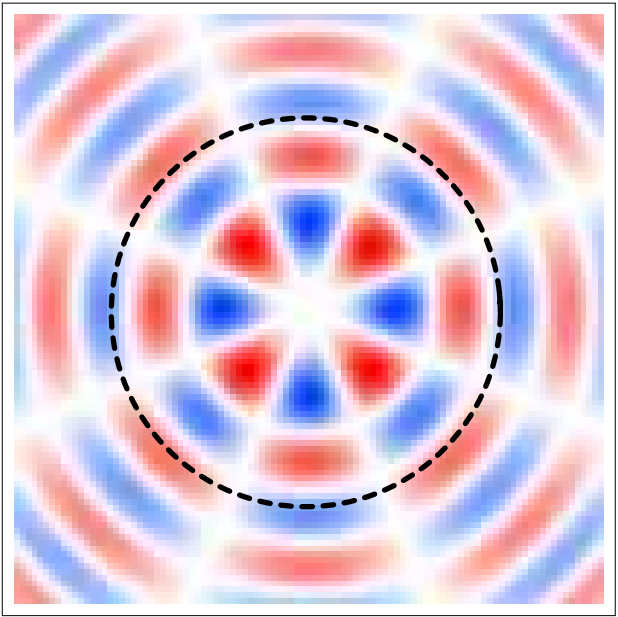}
  \caption{View of the polar cap of pulsar from the top.
    Goldreich-Julian charge density for oscillation mode $(44,4)$ is
    shown by the color map (positive values in red, negative -- in blue).
    The quantity changes harmonically with time. The polar cap
    boundary for a pulsar with period 3 ms is shown by the dashed
    line.  }
\label{fig:Rgj__44_4}
\end{figure}

\begin{figure}
  \centering
  \includegraphics[clip,totalheight=6cm]{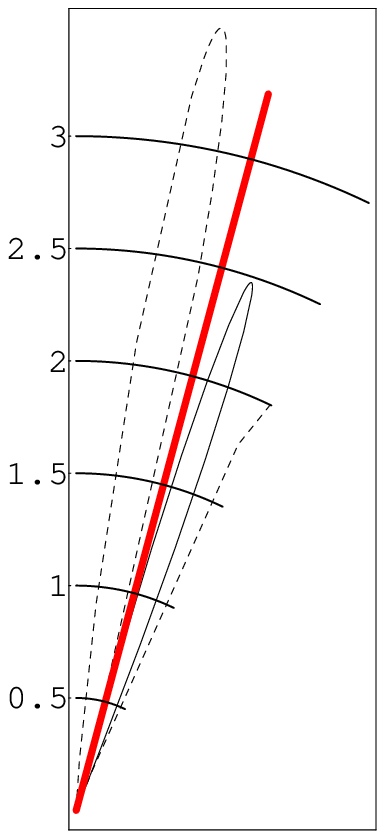}
  \ 
  \includegraphics[clip,totalheight=6cm]{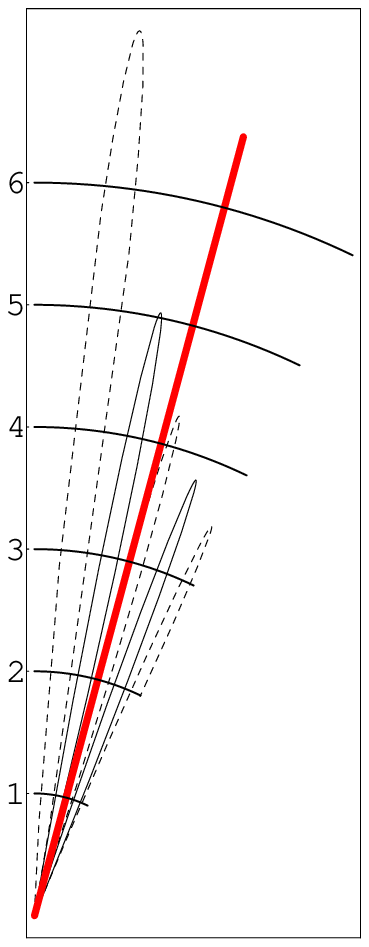} 
  \ 
  \includegraphics[clip,totalheight=6cm]{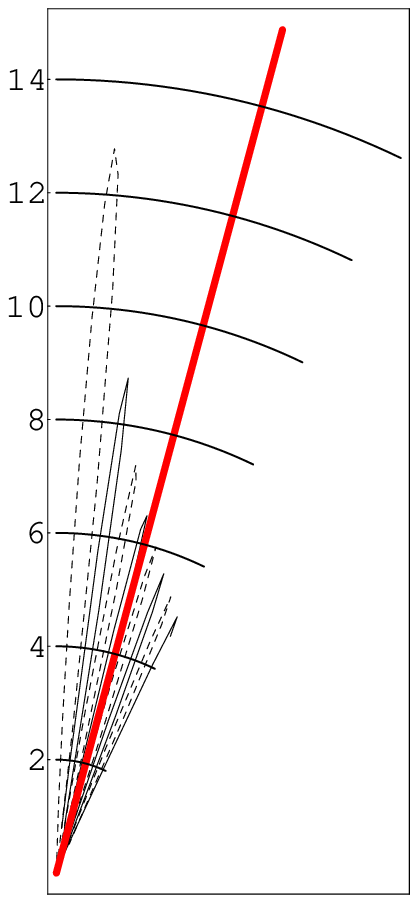} 
  \caption{Dependency of the Goldreich-Julian charge density on the
    harmonic number $l$. Distribution of \Rgj{} on the NS surface in the
    polar cap of pulsar is shown for oscillation modes (from left to
    right) $(28,4)$, $(44,4)$, $(64,4)$ in polar coordinates
    $(\left|\Rgj^{lm}(\RS,\theta,0)\right|, \theta)$. Positive values
    of charge density are shown by the solid line and negative ones by
    the dashed line. The red line shows the angle at which the last
    closed field line intersect the NS surface for a pulsar with
    period 3 ms. Circular segments correspond to the values of \Rgj{}
    in normalized units shown at the left side of each plot. The same
    normalization for \Rgj{} is used as in Fig.~\ref{fig:Rgj_surf}.}
  \label{fig:Rgj_increase_l}
\end{figure}

\begin{figure}
  \centering
  \includegraphics[clip,totalheight=6cm]{fig7_1.eps}
  \ 
  \includegraphics[clip,totalheight=6cm]{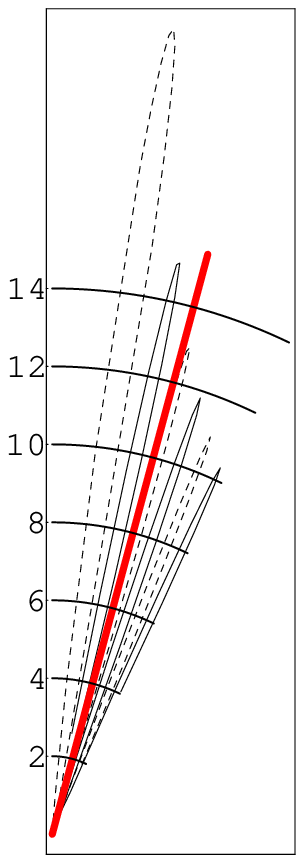}
  \ 
  \includegraphics[clip,totalheight=6cm]{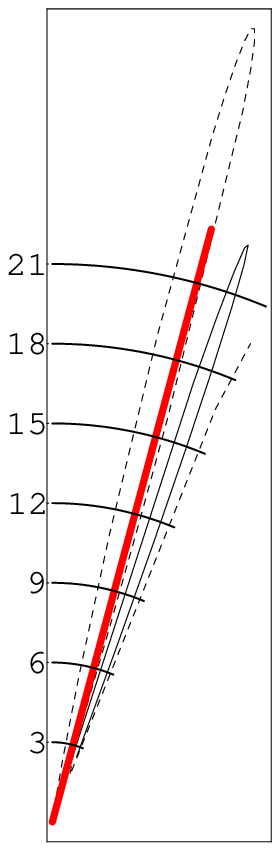} 
  \caption{Dependency of the Goldreich-Julian charge density on the
    harmonic number $m$. Distribution of \Rgj{} on the NS surface in the
    polar cap of pulsar is shown for spheroidal oscillation modes
    (from left to right) $(64,4)$, $(64,8)$, $(64,14)$ in polar
    coordinates $(\left|\Rgj^{lm}(\RS,\theta,0)\right|, \theta)$.
    Meaning of the lines is the same as in
    Fig.\ref{fig:Rgj_increase_l}.}
\label{fig:Rgj_increase_m}
\end{figure}

As expected, the GJ charge density distribution follows the
distribution of the velocity field (see e.g.
Figs.~\ref{fig:Rgj_map},~\ref{fig:Rgj_surf},~\ref{fig:Rgj__44_4}).
With increasing of the harmonic numbers, \Rgj{} falls more rapidly
with the distance, what is also expected for multipolar solutions.  A
remarkable property of GJ charge density distribution near oscillating
star is that the local maxima of \Rgj{} increases with increasing of
both $l$ and $m$ (see
Figs.~\ref{fig:Rgj_increase_l},~\ref{fig:Rgj_increase_m}).  The reasons
for this are as follows. The electric field induced by oscillations is
of the order
\begin{equation}
  \label{eq:Egj_estimate}
  \egj \sim \frac{\Vosc}{c} B\:.
\end{equation}
The charge density supporting this electric field is of the order of
$E/\Delta{}x$, where $\Delta{}x$ is a characteristic distance of
electric field variation.  For a mode with harmonic number $l$ this
size is of the order of $\RS/l$.  Hence, for \Rgj{} we have an
estimate
\begin{equation}
  \label{eq:Rgj_estimate}  
  \Rgj \sim  
  l\:\frac{E}{4\pi\RS} \sim 
  l\:\frac{\Vosc}{c}\:\frac{B}{4\pi\RS}\:.
\end{equation}
So, for the same amplitude of velocity of oscillation the amplitude of
variation of the GJ charge density is larger for higher harmonics.
Equation~(\ref{eq:Rgj_estimate}) is in a good agreement with the exact
results shown in
Figs.~\ref{fig:Rgj_increase_l},~\ref{fig:Rgj_increase_m}.

\subsection{Particle acceleration in the polar cap of pulsar}
\label{sec:part-accel-polar}

The GJ charge density induced by NS oscillations influences particle
acceleration mechanism in the polar cap of pulsar.  As we will show,
oscillations will have the strongest impact on the accelerating
electric field in the polar cap of pulsar for models with free
particle escape from the NS surface
\citep{Arons/Scharlemann/78, Muslimov/Tsygan92}.  The accelerating
electric field in pulsars arises due to deviation of the charge
density of the plasma from the local GJ charge density.  For pulsar
models with Space Charge Limited Flow (SCLF) the charge density of the
flow $\rho$ at the NS surface is equal to the local value of the GJ
charge density, $\rho(\RS)=\Rgj(\RS)$.  Magnetic field lines diverge
and the charge density of the flow decreases with increasing of the
distance from the star.  However, the local GJ charge density
decreases in a different way and at some distance $r$ from the NS
$\Rgj(r)\neq\rho(r)$.  The discrepancy between these charge densities
gives rise to a longitudinal electric field
\citep[see][]{Arons/Scharlemann/78,Muslimov/Tsygan92}.

The GJ charge density for oscillation modes with large $l$, $m$ falls
very rapidly with the distance.  Hence, the charge density of a
charge-separated flow for oscillating NS will exceed the local GJ
charge density at some distance from the star.  This produces a
\emph{decelerating} electric field.  Therefore, in the case of
rotating and oscillating NS, if $\Osc{\Rgj}<\Rotat{\Rgj}$, the
effective accelerating electric field will be reduced periodically due
to superposition of accelerating and decelerating electric fields.

\begin{figure}
  \centering
  \includegraphics[width=.9\columnwidth]{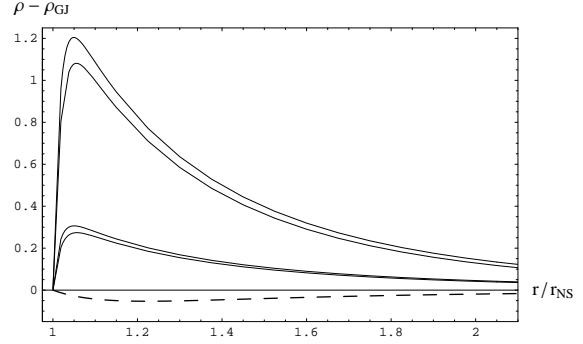}
  \caption{$\Delta\rho\equiv\rho-\Rgj$ -- difference between the charge
    density of a space charge limited flow and the local
    Goldreich-Julian charge density along magnetic field lines in the
    polar cap of pulsar.  $\Delta\rho$ is shown (in arbitrary units)
    by solid lines for spheroidal modes with different ($l,m$), from
    top to bottom: (64,14), (54,14),(64,2),(54,2). The same relation
    for an aligned rotator is shown by the dashed line. Negative
    values of $\Delta\rho$ give rise to an accelerating electric
    field, negative ones -- to a decelerating field.}
\label{fig:R_m_Rgj}
\end{figure}

The oscillational GJ charge density $\Osc{\Rgj}$ for modes with large
$l$, $m$ decreases practically to zero already at 2-3 NS radii and the
whole oscillational GJ charge density contributes to the decelerating
electric field.  While in the case of rotation only $\sim{}15\%$ of
the GJ charge density contributes to the accelerating electric field
for SCLF in the polar cap \citep{Muslimov/Tsygan92}.  The most
important factor increasing modification of the accelerating electric
field is that the amplitude of $\Osc{\Rgj}$ increases with increasing
of the harmonic numbers of the mode.  In Fig.~\ref{fig:R_m_Rgj} we
show the difference between the charge density of SCLF and the local
GJ charge density for dipolar magnetic field as a function of the
distance for rotating and oscillating stars.  In order to demonstrate
the importance of the discussed effects we consider a rather grotesque
case when the linear velocity of rotation is equal to the maximum
velocity of oscillations.  It is evident from this plot, that for
large enough $l$ and $m$ the decelerating electric field caused by
stellar oscillations could be of comparable strength with the
accelerating electric field even if $\Vosc\ll\Vrot$.

Let us estimate the harmonic number of the oscillation mode where
decelerating electric field would have a given impact on the
accelerating electric field induced by NS rotation.  The decelerating
electric will be $\kappa$ times less that the rotational accelerating
electric field,
\begin{equation}
  \kappa \equiv \frac{ \Osc{E}_\mathrm{dec} }{ \Rotat{E}_\mathrm{acc} }
  \sim \frac{ \Osc{\Rgj} }{ 0.15\Rotat{\Rgj} }\:,
  \label{eq:kappa}
\end{equation}
if 
\begin{equation}
  \label{eq:l_Eacc_Edec_V}
  l \sim 0.15\:\kappa\: \frac{\Vrot}{\Vosc}\:,
\end{equation}
here \Vrot{} is the linear velocity of NS rotation at the equator.
For example for $l\sim{}100$, from this equation we can see, that for
canceling of the accelerating electric field it is sufficient that
the velocity amplitude is only $\sim{}10^{-3}$ of rotational velocity
at the NS equator.

For pulsar operating in \citet{Ruderman/Sutherland75} regime the
impact of stellar oscillations on the accelerating field in the polar
cap will be reduced by the factor of $\sim{}10$.  In this model the
accelerating electric field is generated in a vacuum gap, so the whole
rotational GJ charge density contribute to the accelerating electric
field and in this case $\kappa\sim\Osc{\Rgj}/\Rotat{\Rgj}$.

\section{Discussion}
\label{sec:discussion}

We have shown, that oscillations of the NS can induce changes in the
accelerating electric field, which are more stronger than a naive
estimation $(\Vosc/c)B$.  Indeed, for high harmonics the induced
electric field will be $\sim{}l$ times stronger.  In order to make
definitive predictions about observable parameters is necessary to
study the polar cap acceleration zone more detailed.  An accelerating
electric potential and the height of the pair formation front should
be calculated.

However, we can do simple estimations using
equation~(\ref{eq:l_Eacc_Edec_V}).  Let us estimate the harmonic
number of the mode which can cancel the accelerating electric field,
assuming the mode is excited by a glitch.  As we pointed in
Sec.~\ref{sec:general-formalism} only modes with non-zero amplitude on
the NS surface can produce changes in the magnetosphere.  The
distribution of oscillational motion plays crucial role.  If
oscillations are trapped in the NS crust a rather small energy will be
required to pump the oscillation amplitude to the level high enough
for strong disturbance of the accelerating electric field.  If a
fraction $\epsilon$ of the NS mass $\NS{M}$ is involved in the
oscillations the amplitude of the oscillational velocity is of the
order
\begin{equation}
  \label{eq:Vosc_MNS}
  \Vosc \sim \sqrt{ \frac{2\Osc{W}}{\epsilon \NS{M}} }\:,
\end{equation}
where $\Osc{W}$ is the total energy of the mode.  The energy
transferred during the glitch of the amplitude $\Delta\Omega$ is
\begin{equation}
  \label{eq:Wglitch}
  W^\mathrm{glitch} = 
  i\;\NS{I} \Omega \Delta \Omega = 
  i\;\NS{I} \left(\frac{2\pi}{P}\right)^2
  \frac{\Delta\Omega}{\Omega} \:,
\end{equation}
where $i$ is the fraction of the total momentum of inertia of the NS
$\NS{I}$ coupled to the crust.  Let us assume that some fraction
$\eta$ of this energy goes into excitation of oscillations.  Using
eqs.~(\ref{eq:Wglitch}),~(\ref{eq:Vosc_MNS}),~(\ref{eq:l_Eacc_Edec_V})
we get conditions for harmonic number of modes which would
periodically cancel the accelerating electric field, as
\begin{equation}
  \label{eq:l_blok_final_SCLF}
   {\bf l} > 300 \;
   {\eta_\%^{-1/2}} \;
   {\epsilon^{1/2}}\; 
   {i^{-1/2}}\;
  \left( \frac{\Delta\Omega}{\Omega} \right)_6^{-1/2}
  \;,
\end{equation}
for SCLF model, and for \citet{Ruderman/Sutherland75} model:
\begin{equation}
  \label{eq:l_blok_final_RS}
  {\bf l} > 2000 \;
  {\eta_\%^{-1/2}} \;
  {\epsilon^{1/2}}\; 
  {i^{-1/2}}\;
  \left( \frac{\Delta\Omega}{\Omega} \right)_6^{-1/2}
  \;.
\end{equation}
Here $\eta_\%$ is measured in per cents and the relative magnitude of
the glitch $\left(\Delta\Omega/\Omega\right)_6$ is normalized to
$10^{-6}$.  It is widely accepted, that the origin of pulsar glitches
is angular momentum transfer from the NS core to the crust. In the
frame of this model the fraction of the energy which can go into
excitation of NS oscillation $\eta$ is of the order of
$\Delta\Omega/\Omega$, i.e. it is very small, of the order of
$\sim{}10^{-6}$.  We may speculate however, that excited oscillations
are trapped in the NS crust, i.e. $\epsilon$ is also very small. Such
global oscillation modes ($l\sim$ several hundreds) could induce
substantial changes in the accelerating electric field.

As we mentioned in Sec.~\ref{sec:general-formalism} all physical
quantities in the solutions obtained here oscillate with the frequency
of the star oscillations.  The accelerating electric field close to
the local geometrical maxima of the oscillational GJ charge density
will be weakened periodically by the decelerating effect due to
stellar pulsations.  The field oscillation will influence the particle
distribution in the open field line zone of the pulsar magnetosphere
and it should produce some observable effects.  Depending on
oscillation mode and position of the line of sight a complicated
pattern will appear periodically in individual pulse profiles.
Although individual pulses are highly variable, the presence of
periodical features should be possible to discover in the power
spectra of pulsars, provided the oscillations are excited to a high
enough level and observations have been made with hight temporal
resolution.  If one observes some feature, which appears just after the
glitch, then decreases and disappears after some time, and it never
appears in the normal pulsar emission, then one can undoubtedly
attribute this feature to the NS oscillations.

\begin{acknowledgements}
  I wish to thank A.~Alpar for discussion.
  This work was partially supported by RFBR grant 04-02-16720, and by
  the grants N.Sh.-5218.2006.2 and RNP-2.1.1.5940
\end{acknowledgements}

\bibliographystyle{spmpsci}
\bibliography{ns_oscill}   

\end{document}